\documentclass[12pt]{iopart}

\usepackage{iopams}  
\usepackage[breaklinks=true,colorlinks=true,linkcolor=blue,urlcolor=blue,citecolor=blue]{hyperref}
\usepackage{geometry}
\usepackage{graphicx}
\usepackage{caption}
\usepackage{subcaption}
\usepackage{cite}
\usepackage{tabularx}
\usepackage{xfrac}
\begin{document}

\title{Adaptive survival movement strategy to local epidemic outbreaks in cyclic models}

\author{J. Menezes\textsuperscript{1,2}, B. Moura\textsuperscript{3,4} and E. Rangel\textsuperscript{2}}

\address{$^1$ Institute for Biodiversity and Ecosystem
Dynamics, University of Amsterdam, Science Park 904, 1098 XH
Amsterdam, The Netherlands}

\address{$^2$ School of Science and Technology, Federal University of Rio Grande do Norte\\
59072-970, P.O. Box 1524, Natal, RN, Brazil}

\address{$^3$  Edmond and Lily Safra International Institute of Neuroscience, Santos Dumont Institute,
Av Santos Dumont 1560, 59280-000, Macaiba, RN, Brazil} 

\address{$^4$ Department of Biomedical Engineering, Federal University of Rio Grande do Norte,
 Av. Senador Salgado Filho 300, Lagoa Nova, 59078-970, Natal, RN, Brazil}


\ead{jmenezes@ect.ufrn.br}


\begin{abstract}
We study the generalised rock-paper-scissors game with five species whose organisms face local epidemic outbreaks.
As an evolutionary behavioural survival strategy, organisms of one out of the species move in the direction with more enemies of their enemies to benefit from protection against selection. 
We consider that each organism scans the environment, 
performing social distancing instead of agglomerating when perceiving that the density of sick organisms is higher than a tolerable threshold. 
Running stochastic simulations, we study the interference
of the adaptive movement survival strategy in spatial pattern formation, calculating the characteristic length scale of the typical spatial domains inhabited by organisms of each species.
We compute how social distancing trigger impacts 
the chances of an individual being killed in the cyclic game and contaminated by the disease. The outcomes show that the species predominates in the cyclic game because of the organisms' local adaptation. The territory occupied by the species grows with the proportion of individuals learning to trigger the social distancing tactic.
We also show that organisms that perceive large distances more properly execute the adaptive strategy, promptly triggering the social distancing tactic and choosing the correct direction to move.
Our findings may contribute to understanding the role of adaptive behaviour when environmental changes threaten biodiversity.
\end{abstract}

%
%
%
%
%

\section{Introduction}
\label{sec:int}

Spatial interactions among species play a vital role in the stability of ecosystems \cite{ecology,foraging,BUCHHOLZ2007401}. It has been reported that many species evolve to better adapt to the local environmental conditions, increasing the chances of controlling the natural resources to ensure species persistence \cite{climatechange,adap2}. Motivated by local stimuli, many animals move strategically as an adaptive response to alterations in their living conditions \cite{butterfly,adaptive1,adaptive2,Dispersal,BENHAMOU1989375,Causes,MovementProfitable}. Following environmental cues, organisms can escape threats to their existence, making alliances with other species to protect them against natural enemies' attacks. 
Furthermore, scanning the neighbourhood helps individuals observe additional survival hazards ravaging the ecosystem, such as epidemic outbreaks \cite{epidemicbook,epidemicprocess,COVID,disease2,disease4,DONOFRIO2022112072}. 
Using sensory information, individuals can decide when it is decisive to flee from enemies or perform social distancing strategies \cite{Odour,socialdist,soc}.

There is plenty of evidence that self-preservation behavioural strategies play a central role in controlling natural resources. 
Therefore, prioritising moving toward the direction where the probability of being killed is minimum has been demonstrated to be more advantageous in cyclic game models \cite{Moura,enenyenemy,ref10-0,ref10-1,ref10-2}. It has been shown in Ref. ~\cite{Moura} that the Safeguard strategy is advantageous for species participating in a cyclic spatial game, giving the species the possession of the most significant fraction of the territory. 
However, as investigated in Ref.~\cite{combination}, in the event of an epidemic strikes the system, individuals' concern about survival increases: besides being eliminated in the cyclic spatial game, they are vulnerable to being contaminated by the infectious disease that is transmitted person-to-person. 
In this scenario, the Safeguard tactic, which involves a social approach, may become problematic since the organism must be distant from all possible viral vectors. To avoid contamination, thus, organisms perform the Social Distancing strategy, moving toward the direction with the highest density of empty species \cite{combination}.
But neglecting the protection against enemies to perform the Social Distancing strategy is a dilemma that needs to be well considered by each organism. Combining both survival movement strategies provides better protection because if executing only Social Distancing, the individual is vulnerable to the attack of enemies \cite{combination}.

In this work, instead of a fixed global design for the combination of survival movement strategies as investigated in Ref.~\cite{combination}, valid for all organisms in the system, each individual scans the neighbourhood to observe the presence of sick individuals. This allows the individual to decide which survival strategy to perform at each instant, switching the self-preservation strategy in response to a change in the local living conditions. 
Epidemic outbreaks can reach departed regions differently because the pathogen that causes the disease is passed from individual to individual. Thus, each organism needs to be aware of the arrival of new disease waves in the local community. If the chances of being infected are higher than a tolerable threshold, Social Distancing is executed; otherwise, the Safeguard tactic is continued.
The decision is based on the individual's capacity to perceive the neighbourhood: i) how far the organisms can scan their neighbourhood; ii) the cognitive and adaptive ability to interpret the signals received and change from the Safeguard to Social Distancing strategy when facing a local epidemic outbreak.

Our goal is to understand how the adaptive survival strategy to local epidemic outbreaks modifies the pattern formation process and changes the average size of the typical spatial domains. Then, we aim to quantify the benefits of behavioural tactics in the organisms' selection and infection risks. Our simulations consider various tolerable thresholds for organisms to trigger the Social Distancing strategy.
The research is performed using stochastic simulations
based on the May-Leonard implementation of the rock-paper-scissors model, where the total number of individuals is not conserved \cite{ref1-1,ref1-2,Szolnoki-JRSI-11-0735,Anti1,anti2,MENEZES2022101606,PhysRevE.97.032415,Avelino-PRE-86-036112,MENEZES2022111903,ref10-3}. 
The disease spreading happens between two immediate neighbours in the lattice, with a sick individual transmitting the virus to a healthy organism, irrespective of the species \cite{combination,rps-epidemy,epidemic-graphs,germen}. Sick organisms act as viral vectors, with a probability of being cured or dying, according to the disease virulence; no organism or species is immune, even after being cured of the disease.

The outline of this paper is as follows. The methods are described in Sec.~\ref{sec2}, where we introduce the model and the simulation implementation.
Next, we explore the role of the locally adaptive survival movement tactic in the spatial patterns and calculate the dynamics of the densities of species in Sec.~\ref{sec3}.
In Sec.~\ref{sec4}, we compute the impact of the local strategies on organisms' selection and infection risks; the densities of species and the characteristic scale of the typical areas occupied by each species 
in terms of the threshold are also computed. 
The dependence of the outcomes on the individuals' physical abilities to perform the survival strategy is quantified in Sec.~\ref{sec5}. 
Finally, we discuss the results and present our conclusions in Sec.~\ref{sec6}.

\begin{figure}
\centering
\includegraphics[width=50mm]{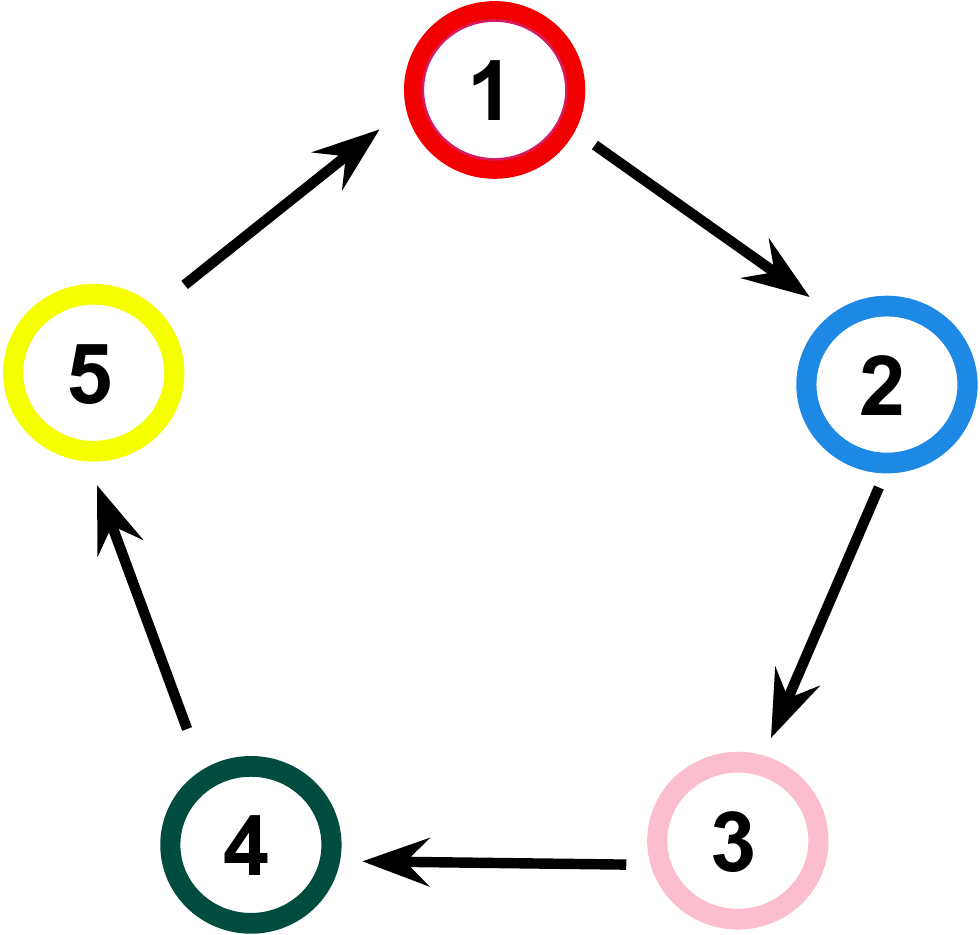}
\caption{Illustration of the generalised rock-paper-scissors model with $5$ species. Selection interactions are represented by arrows indicating the dominance of organisms of species $i$ over individuals of species $i+1$.}
\label{fig1}
\end{figure}


\section{Methods}
\label{sec2}

\subsection{The model}
We investigate a cyclic game system composed of five species whose selection interactions follow the generalised rock-paper-scissors model, 
illustrated in Fig.~\ref{fig1}. Accordingly, individuals of species
$i$ dominate over individuals of species $i+1$, with $i=1,2,3,4,5$; we assume the identification $i=i+5\,\alpha$, where $\alpha$ is an integer. We consider that an epidemic spreads through the system, with organisms of all species equally susceptible to infection and reinfection by a disease 
transmitted person to person. All infected organisms become viral vectors to contaminate their immediate neighbours; some sick individuals die while others are cured.

In our model, organisms of one out of the species perform behavioural movement strategies to protect themselves against nearby enemies and infection by approaching viral vectors. 
Individuals usually perform the Safeguard strategy, moving towards the direction with the highest density of enemies of their enemies \cite{Moura}. Therefore, they profit from being close to guards, decreasing the risk of being killed in the cyclic game. However, if the density of sick individuals in the vicinity represents a danger of imminent contamination, the individual opts for the Social Distancing tactic \cite{combination}. The decision to move to the areas with lower population density is individual, being triggered every time a surge affects the organism's neighbourhood making the local density of sick individuals higher than a given threshold.
\subsection{Simulations}
We use square lattices with periodic boundary conditions to implement our stochastic simulations. As the spatial interactions may lead to organisms' births and deaths, we use the May-Leonard implementation: the total number of individuals is not conserved \cite{leonard}. We assume that each grid point contains at most one individual, which limits the maximum number of individuals to equal the total number of grid points, $\mathcal{N}$. 
The densities of individuals of species $i$ at time $t$ is defined as 
$\rho_i = I_i/\mathcal{N}$, where $I_i$ is number of individuals of species $i$,
with $i=1,2,3,4,5$.

We identify the organisms using the notation $h_i$ and $s_i$ for
healthy and sick individuals; the use of the identification $i$ stands for individuals irrespective of illness or health, with $i=1,2,3,4,5$. This notation allows the description of the spatial interactions as \cite{combination}:
\begin{itemize}
\item
Selection: $ i\ j \to i\ \otimes\,$, with $ j = i+1$, where $\otimes$ means an empty space; selection interactions happens between individuals, irrespective of the healthy conditions;
\item
Reproduction: $ i\ \otimes \to i\ h_i\,$; a reproduction interaction generates a new healthy individual.
\item
Mobility: $ i\ \odot \to \odot\ i\,$, where $\odot$ means either an individual of any species or an empty space; when moving, a healthy or a sick individual switches position with another organism of any species or with an empty site;
\item
Infection: $ s_i\ h_j \to s_i\ s_j\,$, with $i, j=1,2,3,4,5$: a sick individual passes the disease to another organism of any species; 
\item
Cure: $ s_i \to h_i\,$: an ill individual is naturally cured of the disease without gaining immunity, thus being subject to reinfection;
\item
Death: $ s_i \to \otimes\,$: when a sick organism dies because of the disease, its position is left empty.
\end{itemize}
Empty spaces are created when individuals are selected, or sick organisms die because of the disease severity. 

The spatial interactions are executed by assuming the von Neumann neighbourhood: individuals may interact with one of their four immediate neighbours. At each time step, one of the possible implementations is chosen aleatory, according to the respective interaction probabilities: i) healthy individuals: $s_h$ (selection), $r_h$ (reproduction), $m_h$ (mobility); ii) sick organisms: $s_s$ (selection), $r_s$ (reproduction), $m_s$ (mobility), $\omega$ (infection), $c$ (cure), and $d$ (death by disease complications). The parameters are the same for all organisms of every species. 
Our algorithm follows three steps: i) randomly selecting an active individual; ii) using the model probabilities to choose one interaction to be executed; iii) drawing one of the four nearest neighbours to suffer the action, except for the directional movement strategy, where the organism chooses the direction to move according to the specific survival strategy. If the interaction is implemented, one timestep is counted. Otherwise, the three steps are redone. We assume the time unit called generation, defined as the necessary time to $\mathcal{N}$ timesteps to occur.

\subsection{Adaptive survival strategy to local epidemic outbreaks}
We investigate a locally adaptive self-preservation strategy of individuals that use environmental cues to move cleverly according to the current survival conditions.
This means that each organism interprets the signals received from the neighbourhood to identify the existence of local disease surges, thus maximising the protection from disease infection and enemies' attacks.

Each organism usually moves to minimise the chances of being caught and eliminated in the spatial competition game, performing the Safeguard strategy - this is independent of the organism's geographic position \cite{Moura,combination}.
When performing the Safeguard strategy, an individual of species $i$ moves in the direction with the highest concentration of individuals of species $i-2$. This provides protection against individuals of species $i-1$, according to Fig.~\ref{fig1}.
However, the Safeguard tactic increases the chances of contamination during a local epidemic since the virus is transmitted from individual to individual. Because of this, organisms adapt to respond to the arrival of a local disease surge, temporarily switching from the Safeguard to the Social Distancing strategy, moving towards the direction with more empty spaces.

Every time an organism decides to perform the Social Distancing instead of Safeguard strategy, the vulnerability to being killed in the spatial game increases. Therefore, each organism scans the vicinity to conclude how to move. This means that the choice of the survival strategy is local, based on each organism's reality. 
We define a real parameter $\beta$, with $0\,\leq\,\beta\,\leq\,1$, to define the social distancing trigger parameter: a threshold assumed by the organisms as the minimum local density of sick individuals to perform the Social Distancing strategy.
Only if the local density of sick individuals is higher than $\beta$, the Social Distancing is performed. Otherwise, the Safeguard tactic is executed.

\subsection{The adaptation factor}
Adapting the survival movement strategy also depends on the 
organisms' conditioning, meaning their cognitive ability to learn 
how to react to the presence of sick individuals.
We then introduce the adaptation factor, $\lambda$, a real parameter, $0 \leq \lambda \leq 1$,
which indicates the organism's conditioning to respond to a local surge appropriately. For $\lambda=0$, the organism cannot identify the presence of sick individuals, performing only the Safeguard strategy. In the limit $\lambda=1$, the organism is fully conditioned to react to the local surges correctly.

Every time an individual of species $1$ is stochastically chosen to move, the adaptation factor defines the probability of the survival movement strategy being implemented. Namely,
$\lambda$ is defined as the probability of the organism being able to execute the behavioural strategy; $1-\lambda$ is the probability of moving randomly.

\subsection{Implementation of the survival movement strategies}

The numerical implementation of the strategic movement follows the steps \cite{MENEZES2022101606,combination}:
\begin{enumerate} 
\item
We define the perception radius $R$, measured in lattice spacing, as the maximum distance an individual can scan the environmental cues to use the information to decide what strategic movement is suitable at the moment.
\item
We implement a circular area of radius $\mathcal{R}$ centred in the active individual, defining the total grid points the organism is able to analyse.
\item
We calculate the density of sick organisms within the 
disc of radius $R$ surrounding the active individual. In case of the local density is below the threshold $\beta$, the organism decides to perform the Safeguard strategy; otherwise, the Social Distancing tactic is chosen.
\item
We separate the organism's perception area
into four circular sectors in the directions of the nearest neighbour. Thus, we count the number of empty spaces and organisms within each circular sector; individuals on the borders are assumed to be in both circular sectors.
\item
We define the movement direction according to the organism's strategy choice: the direction with the larger number of organisms of species $i-2$ ($h_{i-2}+s_{i-2}$) for Safeguard or the direction with 
more empty spaces for Social Distancing. If more than one direction is equally attractive, a draw is done.
\end{enumerate}

\subsection{Spatial Autocorrelation Function}

To study the effects of the adaptive survival strategy to local epidemic outbreaks in the spatial patterns, we calculate the characteristic length scale of the typical spatial domains occupied by each species. For this purpose, we compute the spatial autocorrelation function $C_i(r)$, with $i=1,2,3,4,5$, in terms of radial coordinate $r$. Defining
the function $\phi_i(\vec{r})$ to describe the position $\vec{r}$ in the lattice occupied by individuals of species $i$, we use the mean value $\langle\phi_i\rangle$ to find the Fourier transform
\begin{equation}
\varphi_i(\vec{\kappa}) = \mathcal{F}\,\{\phi_i(\vec{r})-\langle\phi_i\rangle\}, 
\end{equation}
and the spectral densities
\begin{equation}
S_i(\vec{k}) = \sum_{k_x, k_y}\,\varphi_i(\vec{\kappa}).
\end{equation}

The normalised inverse Fourier transform allows us to find the autocorrelation function for species $i$ as
\begin{equation}
C_i(\vec{r}') = \frac{\mathcal{F}^{-1}\{S_i(\vec{k})\}}{C(0)},
\end{equation}
which we write as a function of $r$ as
\begin{equation}
C_i(r') = \sum_{|\vec{r}'|=x+y} \frac{C_i(\vec{r}')}{min\left[2N-(x+y+1), (x+y+1)\right]}.
\end{equation}
Defining the threshold $C_i(l_i)=0.15$, we define 
the characteristic length scale for spatial domains of species $i$ as $l_i$, with $i=1,2,3,4,5$. 

\subsection{Selection and Infection Risks}

The impact of the locally adaptive organisms' response to their survival is computed by employing two death risks:
a) Selection Risk, $\zeta_i(t)$: the probability of an organism of species $i$ being killed by an individual of species $i-1$ at time $t$; b) Infection Risk, $\chi_i(t)$: the probability of a healthy organism of species $i$ being infected by an ill individual of any species at time $t$.

The numerical implementation of $\zeta_i(t)$ and $\chi_i(t)$ follows the algorithm:
\begin{enumerate}
\item
We count the total number of healthy and sick individuals of species $i$ when each generation commences; 
\item
We calculate how many individuals of species $i$ are selected (killed by individuals of species $i-1$) during the generation; 
\item
We compute the number of healthy organisms infected during the generation;
\item
The selection risk is the ratio between the number of selected individuals and the total number of organisms;
\item 
The infection risk is the ratio between the number of infected individuals and the total number of healthy organisms.
\end{enumerate}
\begin{figure*}
\centering
    \begin{subfigure}{0.24\textwidth}
        \centering
        \includegraphics[width=33mm]{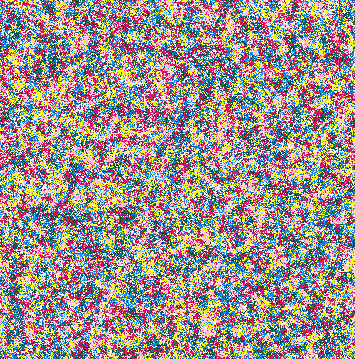}
        \caption{}\label{fig2a}
    \end{subfigure} %
    \begin{subfigure}{0.24\textwidth}
        \centering
        \includegraphics[width=33mm]{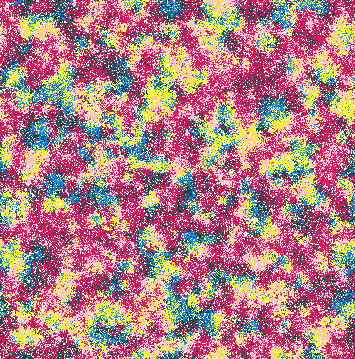}
        \caption{}\label{fig2b}
    \end{subfigure} %
       \begin{subfigure}{0.24\textwidth}
        \centering
        \includegraphics[width=33mm]{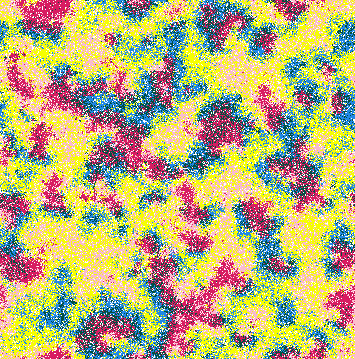}
        \caption{}\label{fig2c}
    \end{subfigure} %
           \begin{subfigure}{0.24\textwidth}
        \centering
        \includegraphics[width=33mm]{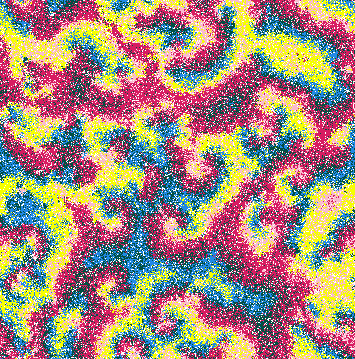}
        \caption{}\label{fig2d}
    \end{subfigure} %
               \begin{subfigure}{0.24\textwidth}
        \centering
        \includegraphics[width=33mm]{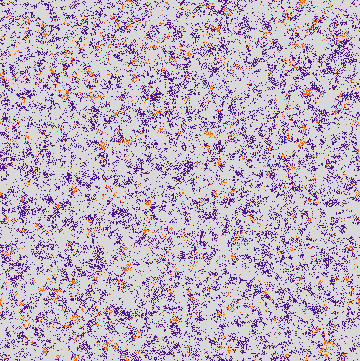}
        \caption{}\label{fig2e}
    \end{subfigure} %
   \begin{subfigure}{0.24\textwidth}
        \centering
        \includegraphics[width=33mm]{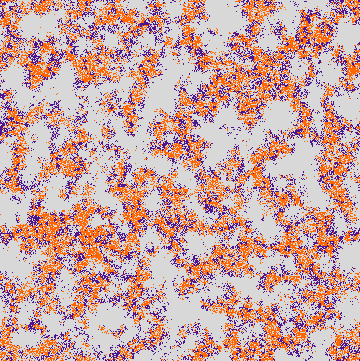}
        \caption{}\label{fig2f}
    \end{subfigure} %
    \begin{subfigure}{0.24\textwidth}
        \centering
        \includegraphics[width=33mm]{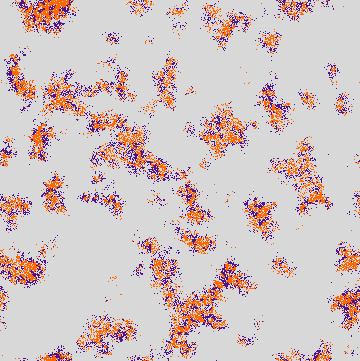}
        \caption{}\label{fig2g}
    \end{subfigure} %
       \begin{subfigure}{0.24\textwidth}
        \centering
        \includegraphics[width=33mm]{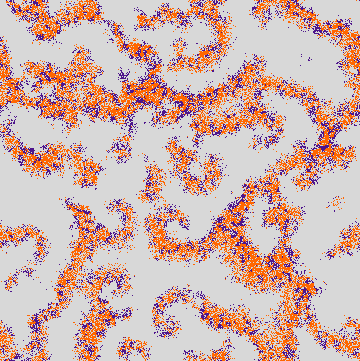}
        \caption{}\label{fig2h}
    \end{subfigure} %
    \caption{
Spatial patterns captured from simulations in lattices with $500^2$ grid points of the generalised rock-paper-scissors game. Figures ~\ref{fig2a}, ~\ref{fig2b}, ~\ref{fig2c}, and ~\ref{fig2d} show snapshots of the spatial configuration after $20$, $100$, $250$, and $2000$ generations. The colours follows the scheme in Fig.~\ref{fig1}; empty spaces appear in white dots. 
Figures ~\ref{fig2e}, ~\ref{fig2f}, ~\ref{fig2g}, and ~\ref{fig2h} highlight the organisms of species $1$, with orange and purples indicating individuals performing Social Distancing and Safeguard tactics, respectively. The social distancing trigger is $\beta=0.6$ and the perception radius is $R=3$. The dynamics of the spatial patterns during the entire simulation are shown in videos https://youtu.be/tRxooHOroZo and https://youtu.be/FYH24c7A1Uc.
}
  \label{fig2}
\end{figure*}

\begin{figure}
\centering
\includegraphics[width=93mm]{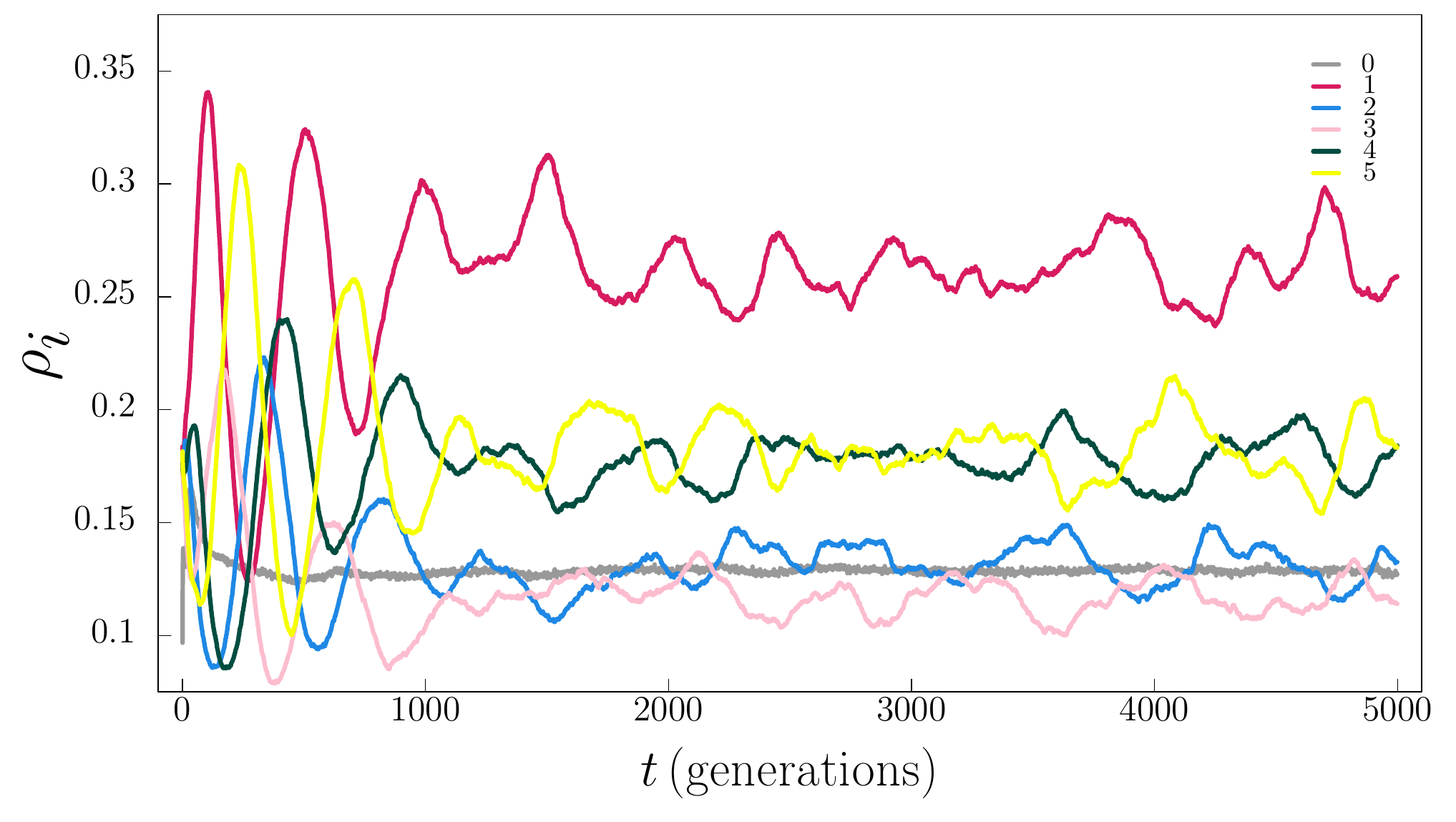}
\caption{Temporal variation of the densities of organisms of species $i$ in the realisation in Fig.~\ref{fig2}. The grey line depicts the variation of the density of empty space, 
while red, blue, pink, green, and yellow stand for species $1$, $2$, $3$, $4$, and $5$, respectively.}
\label{fig3}
\end{figure}

\subsection{Initial conditions, parameters and figures}

Throughout this paper, all outcomes presented were obtained in simulations with $\mathcal{N}=500^2$ grid points, starting from random initial conditions and running for a timespan of $5000$ generations. 
The initial conditions are elaborated by allocating an individual of a random species at each point.
The number of individuals of each species is the same at the beginning of the simulation, $I_i=\mathcal{N}/5$, with $1\%$ being sick - there are no empty spaces in the initial state.
All simulations ran for 
the set of probabilities: $s_h = r_h=m_h=1/3$, $s_s=r_s= 5/31$, $m_s=9/31$, $\omega = 10/31$, and $c=d=1/31$. However, having repeated the simulations for other parameters, we verified that our conclusions are valid for other sets of interaction probabilities.

\begin{figure}
\centering
\includegraphics[width=93mm]{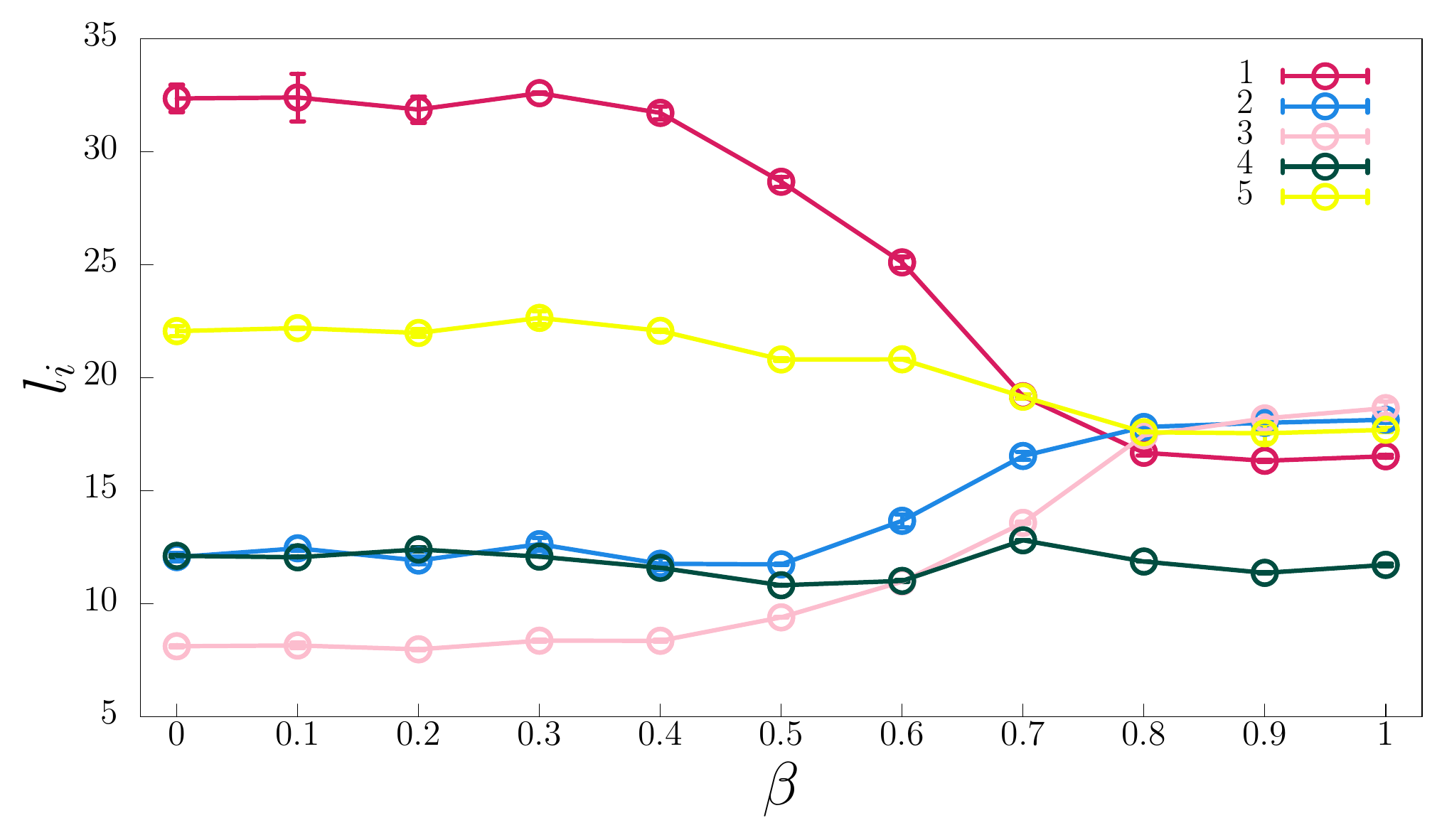}
\caption{Characteristic length scales of the typical spatial domains occupied for species $i$ in terms of the social distancing trigger parameter. 
The results were averaged from sets of $100$ simulations; the error bars indicate the standard deviation. The colours follow the scheme in Fig.~\ref{fig1}.}
\label{fig4}
\end{figure}

\begin{figure}[t]
\centering
    \begin{subfigure}{.47\textwidth}
        \centering
        \includegraphics[width=73mm]{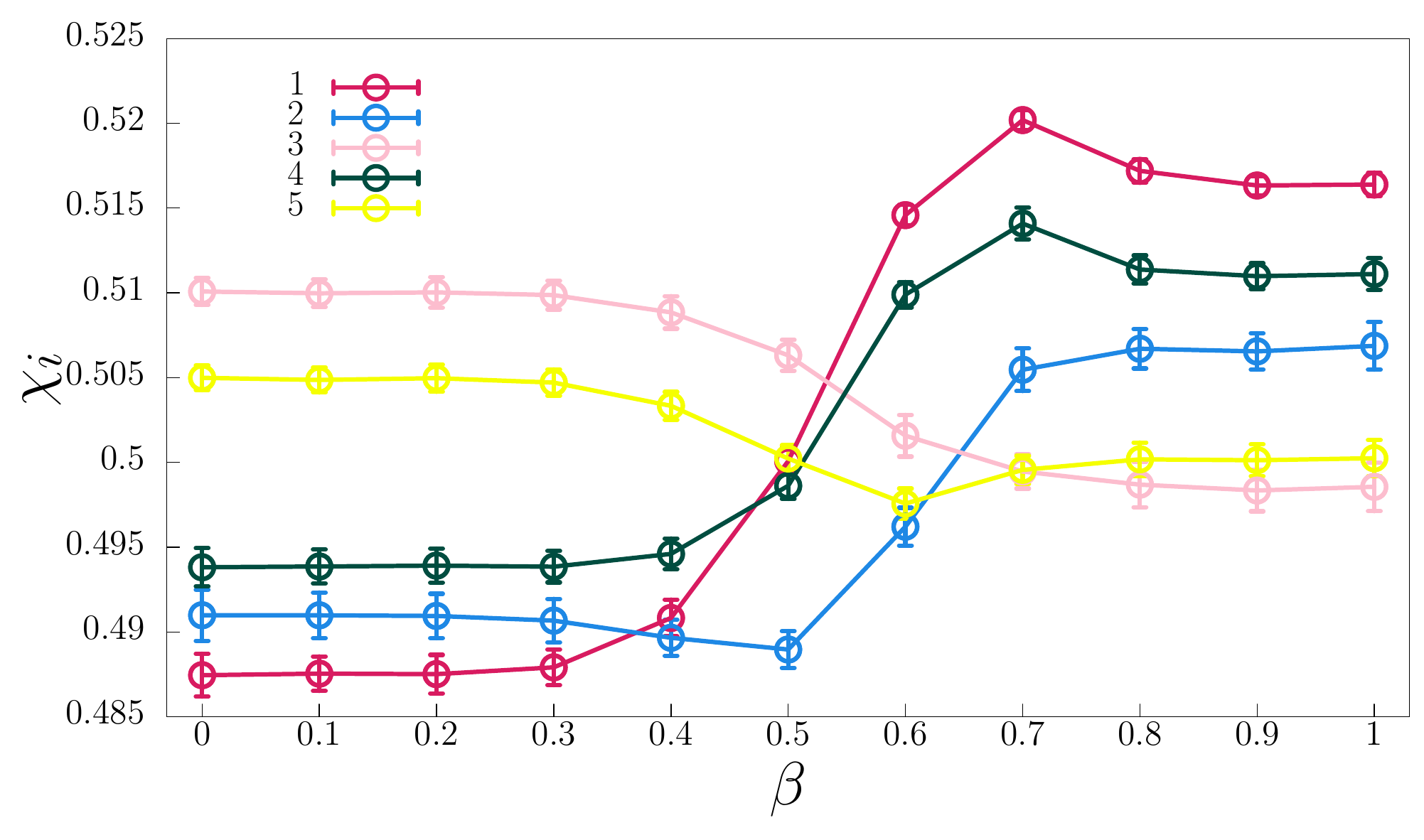}
        \caption{}\label{fig5a}
    \end{subfigure} %
       \begin{subfigure}{.47\textwidth}
        \centering
        \includegraphics[width=73mm]{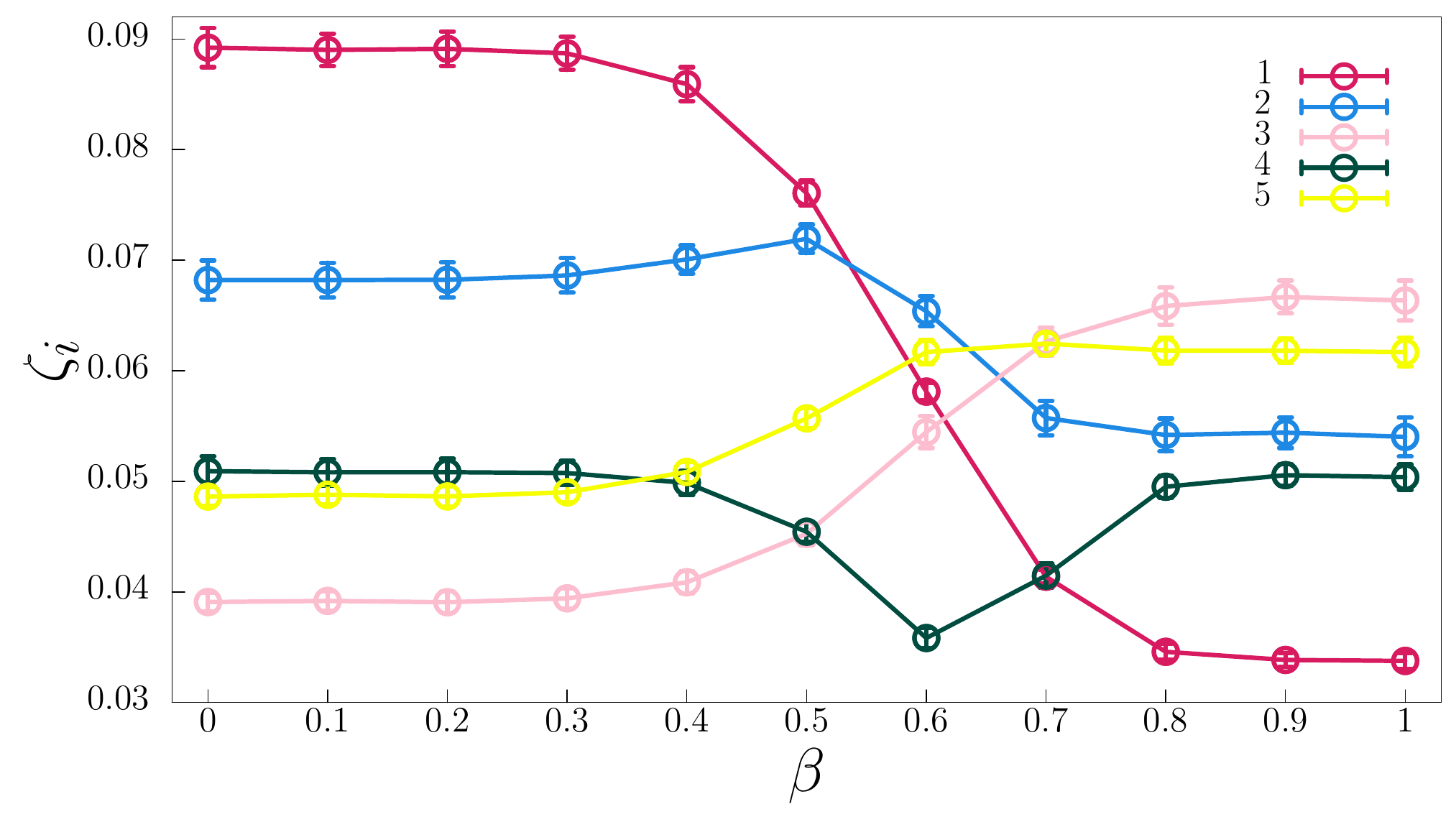}
        \caption{}\label{fig5b}
    \end{subfigure} \\%
           \begin{subfigure}{.47\textwidth}
        \centering
        \includegraphics[width=73mm]{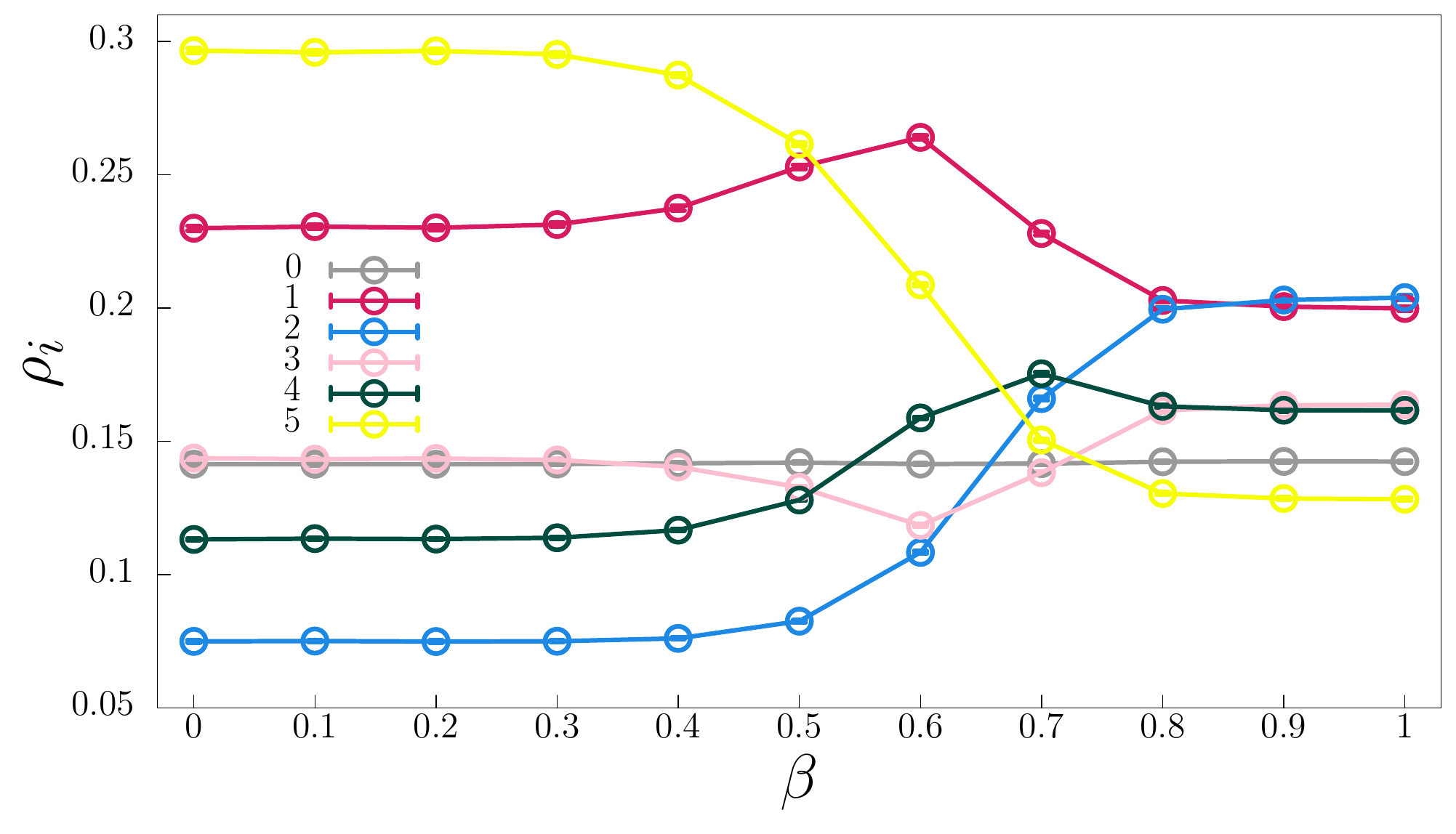}
        \caption{}\label{fig5c}
    \end{subfigure} %
    \caption{Infection risk (Figure a), selection risk (Figure b), densities of species (Figure c) in terms of the social distancing trigger parameter.
The outcomes were averaged from sets of $100$ simulations; the error bars show the standard deviation.}
  \label{fig5}
\end{figure}

The outcomes presented in Fig.~\ref{fig2} and \ref{fig3} were obtained for $R=3$, $\lambda=1.0$, and $\beta=0.6$.
The statistical analyses whose results appear in Figs.~\ref{fig4} to ~\ref{fig7} were realised by performing a series of $100$ simulations, starting from different initial conditions; in each case, the standard deviation is represented by error bars. The parameters were assumed as follows: i) Figs.~\ref{fig4} and ~\ref{fig5}: $0 \leq \beta \leq 1$, with intervals of $\Delta \beta = 0.1$ (for fixed $R=3$ and $\lambda=1.0$); ii) Fig.~\ref{fig6}: $0 \leq R \leq 5$, with intervals of $\Delta R = 1$ (for fixed $\beta=0.6$ and $\lambda=1.0$); iii) Fig.~\ref{fig7}: $0 \leq \lambda \leq 1$, with intervals of $\Delta \lambda = 0.1$ (for fixed $R=3$ and $\beta=0.6$). 

In Figs.~\ref{fig2a} to \ref{fig2d}, organisms of species $i$ are 
depicted by the respective species colours, following the scheme in Fig.~\ref{fig1}, where red, blue, pink, green, and yellow stand for species $1$, $2$, $3$, $4$, and $5$, respectively; white dots show the empty sites. The organisms of species $1$ present in Figs.~\ref{fig2a} to \ref{fig2d} are shown in Figd.~\ref{fig2e} to \ref{fig2h}, with
purple and orange dots indicating the individuals performing Safeguard and Social Distancing, respectively; grey areas are empty or occupied by organisms of species $2$, $3$, $4$, $5$.
The colours in Figs.~\ref{fig3} to ~\ref{fig7} also follow the scheme in Fig.~\ref{fig1}; the grey lines represents empty spaces.

\section{Spatial patterns}
\label{sec3}

To compute the effects of the adaptive survival strategy to local epidemic surges on spatial patterns, we first ran a single simulation considering that organisms react to sick organisms' presence in case the local density of ill individuals is higher than the $60\%$ ($\beta=0.6$). Snapshots of the individuals' spatial distribution after $20$, $100$, $250$, and $2000$ generations are shown in Figs. ~\ref{fig2a}, ~\ref{fig2b}, ~\ref{fig2c}, and ~\ref{fig2d}, respectively. White dots represent empty spaces, whereas organisms are depicted with the respective species' colour in Fig.~\ref{fig1}. The spatial configuration during the whole simulation is shown in the video https://youtu.be/tRxooHOroZo, whereas the temporal dynamics of the densities of species are depicted in Fig.~\ref{fig3}. 

As soon as the simulation commences, organisms of species $1$ cleverly move toward the direction where the death risk is minimum, while all individuals of the other species walk randomly. The advantage of species is reflected in rapid population growth, as observed in the high magnitude in the density of species $1$ in Fig.~\ref{fig3}; the areas with red dots dominate the snapshot in Fig.~\ref{fig2b} ($t=100$ generations).
The high abundance in the density of species $1$ favours organisms of species $5$, that take control of the areas of species $1$ and grow, as shown by the yellow line in Fig.~\ref{fig3} ($t=250$ generations). After the initial stage of high fluctuations in the densities of species (the pattern formation process), spiral waves spread on the lattice. The results show that the cyclic dominance symmetry inherent in the rock-paper-scissors model is broken because deaths of organisms of species $1$ are rarer than fatalities that affect other species. Cleverly adapting the survival movement strategy, organisms guarantee the control of the largest fraction of territory, leading the species $1$ to predominate in the cyclic game, as shown in Fig.~\ref{fig3}.

In addition, Figs. ~\ref{fig2e}, ~\ref{fig2f}, ~\ref{fig2g}, and ~\ref{fig2h} highlight the spatial distribution of organisms of species $1$ in Figs. ~\ref{fig2a}, ~\ref{fig2b}, ~\ref{fig2c}, and ~\ref{fig2d}, respectively. 
When the snapshots were captured, organisms shown in orange were performing the Social Distancing tactic, while the individuals depicted by purple dots were executing the Safeguard tactic. Grey areas are occupied by individuals of species $2$, $3$, $4$, $5$, and empty spaces. Video https://youtu.be/FYH24c7A1Uc shows the dynamics of the organisms of species $1$ executing each survival strategy during the entire simulation.

The results highlight the ability of organisms of species $1$ to 
adjust the survival directional tactic in response to local 
disease surges. 
After an initial proportion of sick individuals is $1\%$, the disease spreads through the lattice, causing local outbreaks. Because of this, the proportion of individuals performing the Social Distancing tactic (orange dots) grows significantly in the first stage of the simulations, as shown in Figs.~\ref{fig2e} and \ref{fig2f}.
However, as each organism performs the Social Distancing strategy only if more than $60\%$ of its neighbours are 
infected, many of them continue executing the Safeguard strategy 
(purple dots). This guarantees the species to profit with the protection against selection and virus infection.
Therefore, changing the directional movement responding to localised concentrations of infected organisms benefits the species whose population grows, as depicted in Fig.\ref{fig3}. 

\section{The role of the social distancing trigger parameter}
\label{sec4}

\subsection{Characteristic length scales of typical spatial domains}

The spatial patterns observed in Fig.~\ref{fig2} show that the asymmetry in the
species segregation results from the local interactions where each organism of species $1$ moves strategically, optimising the results of the behavioural survival strategy according to its own reality. To quantify the characteristic length scale of the typical spatial areas occupied by each species, we ran a set of $100$ simulations for various values of $\beta$. The mean value of characteristic length scale $l_i$, with $i=1,2,3,4,5$ is depicted in Fig.~\ref{fig4}; the error bars represent the standard deviation.

If organisms of species $1$ exclusively execute the Social Distancing tactic, independent of the perception of the local density of sick individuals ($\beta=0.0$), the average size of the areas occupied by them is larger than the others. In this limit case, individuals of species $5$ form the second-largest groups; the characteristic length scale of regions dominated by species $3$ is the shortest. The results are approximately the same for $0.0\,\leq\,\beta\,\leq\,0.3$. However, as the social distancing trigger grows (individuals live with less concern about the disease contamination), the difference in the average sizes of the regions of each species decreases. For $\beta \geq 0.8$, organisms of species $4$ are
limited to the smallest areas; species $1$ occupies spatial domains with the second shortest characteristic length scale.

\subsection{Selection, infection risks, and densities of species}

Performing a series of simulations, starting from different initial conditions, we computed the social distancing trigger parameter
interferes with the organisms' infection, selection risks, and densities of species.
Figures \ref{fig5a}, \ref{fig5b}, and \ref{fig5c} depict $\zeta_i$, $\chi_i$, and $\rho_i$ as functions of $\beta$ - the simulations parameters are the same as in the previous sections. 

Overall, the results show a crossover between scenarios where organisms perform exclusively one out of two survival movement strategies: the Social Distancing ($\beta \leq 0.3$) and Safeguard ($\beta \geq 0.9$) tactics.

\begin{itemize}
\item
If the social distancing trigger is low, the organisms are more careful against disease contamination. For $\beta\,\leq\,0.3$, the fear of being sick is too high that all organisms forget the risk of being killed by enemies for executing exclusively social distancing, as shown in Fig.~\ref{fig5a}.
The consequence is that the infection risk reaches the minimum value while the selection risk is maximum for $\beta\,\leq\,0.3$: the more careful the organisms are against viral infection, the more exposed to being killed by individuals of species $5$. 
\item
Conversely, as $\beta$ grows, the organisms are more tolerant of the presence of viral vectors, concentrating more on approaching the enemies of their enemies. Thus, the selection risk decreases while infection risk grows for larger $\beta$. According to Fig.~\ref{fig5b}, for $\beta \geq 0.9$, the individuals' negligence in avoiding disease contamination leads to the maximum infection risk. This happens because although the protection received from organisms of species $4$ against individuals of species $5$ is maximum, 
there is a chance of part of the guards being sick, thus, infecting the individuals of species $1$. 
\end{itemize}
The results depicted in Fig.~\ref{fig5c} show that the benefits of the adaptive survival strategy for species $1$ are maximised for $\beta=0.6$. In this case, species $1$ predominates in the cyclic game, with the density of organisms of species $1$ reaching the maximum.

\section{The influence of the perception radius}
\label{sec5}
\begin{figure}[t]
\centering
    \begin{subfigure}{.47\textwidth}
        \centering
        \includegraphics[width=73mm]{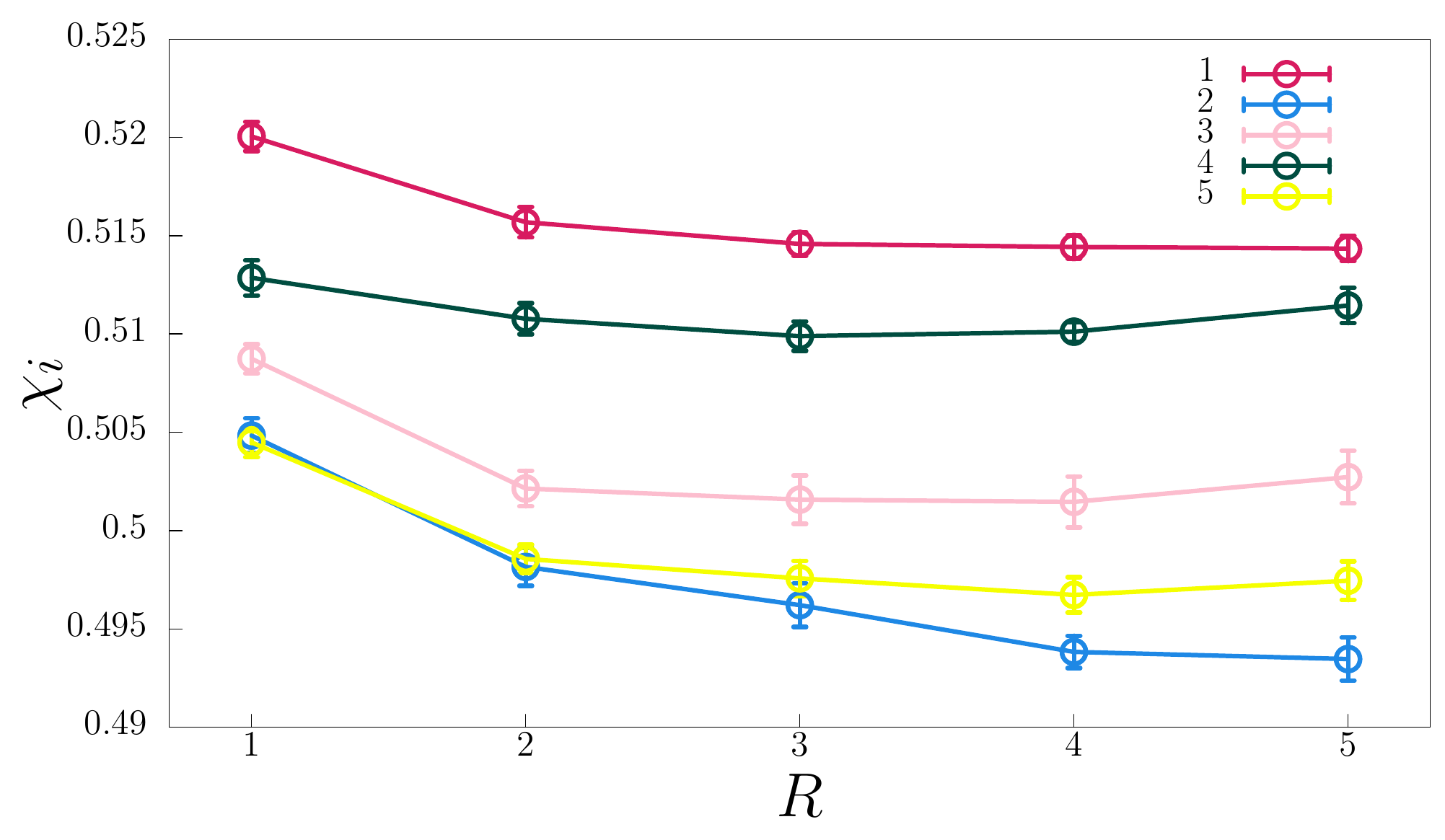}
        \caption{}\label{fig6a}
    \end{subfigure} %
       \begin{subfigure}{.47\textwidth}
        \centering
        \includegraphics[width=73mm]{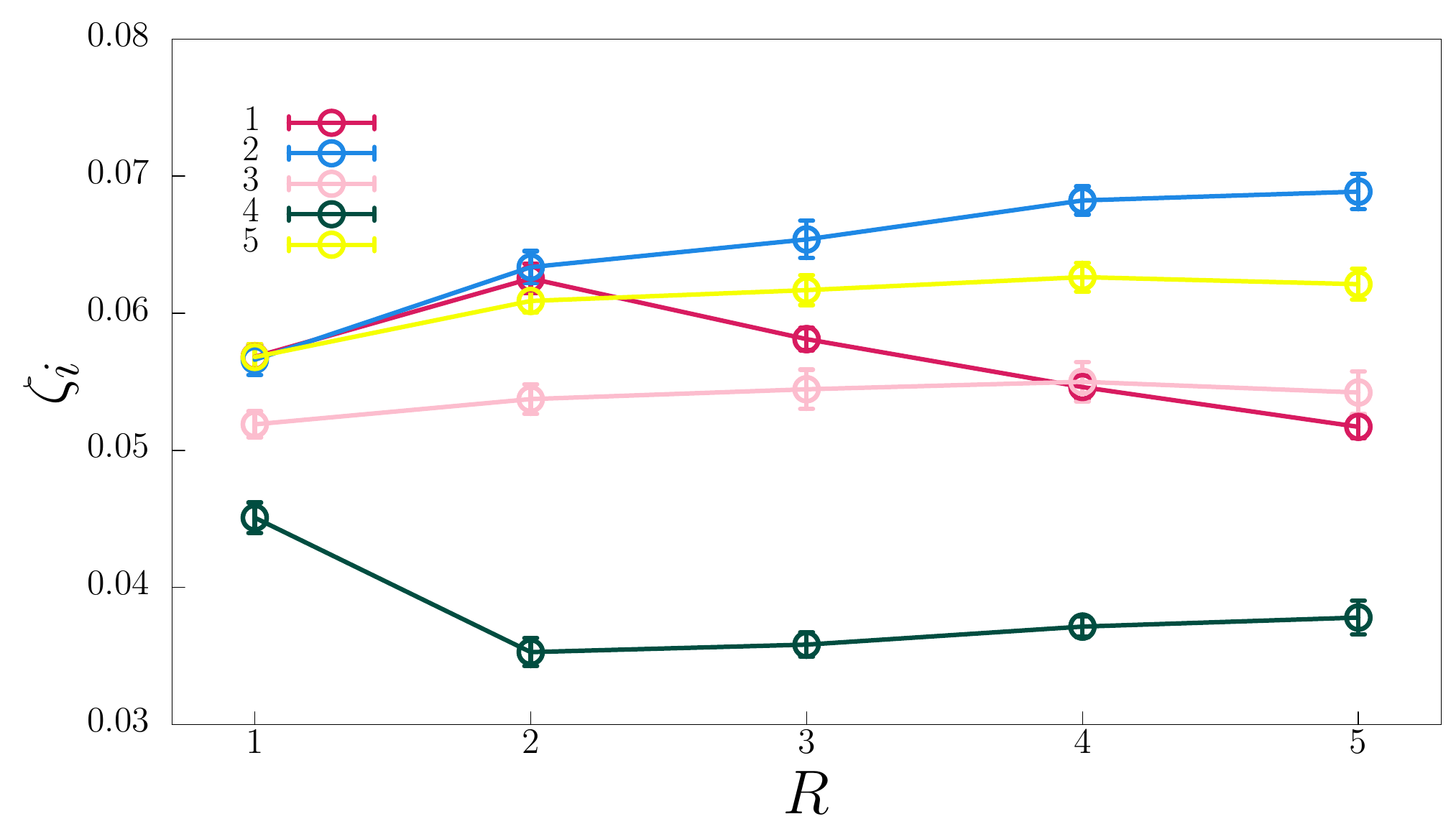}
        \caption{}\label{fig6b}
    \end{subfigure} \\
    \centering
           \begin{subfigure}{.47\textwidth}
        \centering
        \includegraphics[width=73mm]{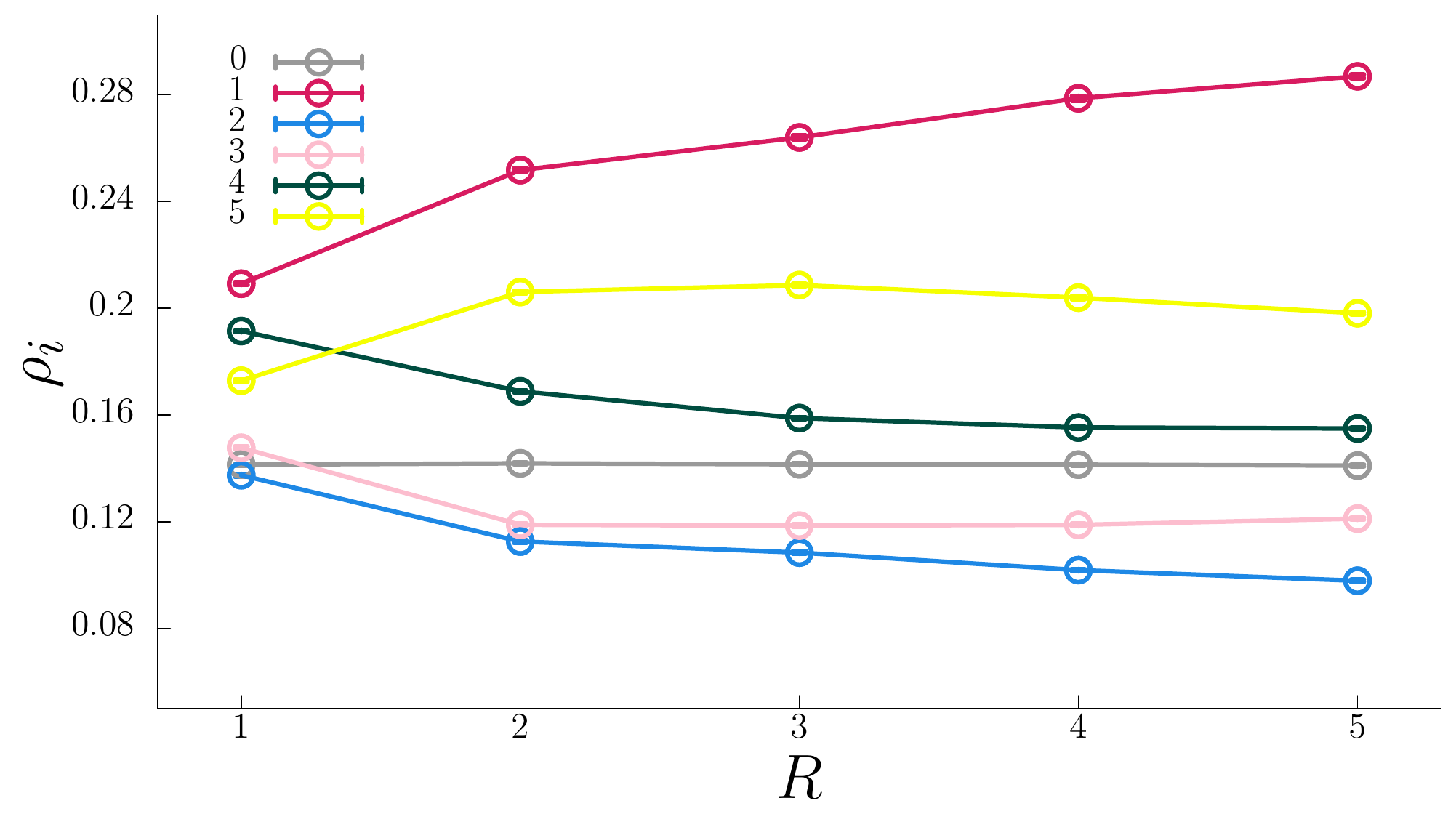}
        \caption{}\label{fig6c}
    \end{subfigure} %
    \caption{Infection risk (Figure a), selection risk (Figure b), and densities of species (Figure c) as functions of the organisms' perception radius.
The results were averaged from $100$ simulations for $\beta=0.6$ and $\lambda=1.0$; the error bars indicate the standard deviation.}
  \label{fig6}
\end{figure}

To investigate the role of the organisms' physical ability to scan the neighbourhood, we performed simulations for various values of $R$ considering $\beta=0.6$ and $\lambda=1.0$, which brings the most significant benefit in terms of population growth for species $1$, as depicted in Fig.~\ref{fig5c}.
Figures \ref{fig6a} and \ref{fig6b} show the organisms' infection and infection risks, respectively; the average densities of individuals of each species appear in Fig.~\ref{fig6c}.

We found that, for short-range perception,
the signals captured provide imprecise information about the neighbourhood. 
Therefore, organisms neither properly detect the danger of contamination by nearby ill organisms nor accurately choose the direction to walk when performing the directional movement strategy. However, even for $R=1$, the locally adaptive strategy is advantageous for species $1$, which dominates in the cyclic game, as depicted in Fig.~\ref{fig6c}. 

As $R$ increases, the environmental cues improve, giving the organisms more precise knowledge of the existence of a local epidemic outbreak: organisms' reaction is more accurate, correctly switching the movement strategies to maximise protection against disease contamination and enemies. The results show that for $R \geq 2$, the reduction in the infection and selection risks accentuates, as depicted in Figs.~\ref{fig6a} and Fig.~\ref{fig6b}. As the number of deaths of individuals of species $1$ decreases for larger $R$, the population rises, as shown in Fig.~\ref{fig6c}.

\begin{figure}[t]
\centering
    \begin{subfigure}{.47\textwidth}
        \centering
        \includegraphics[width=73mm]{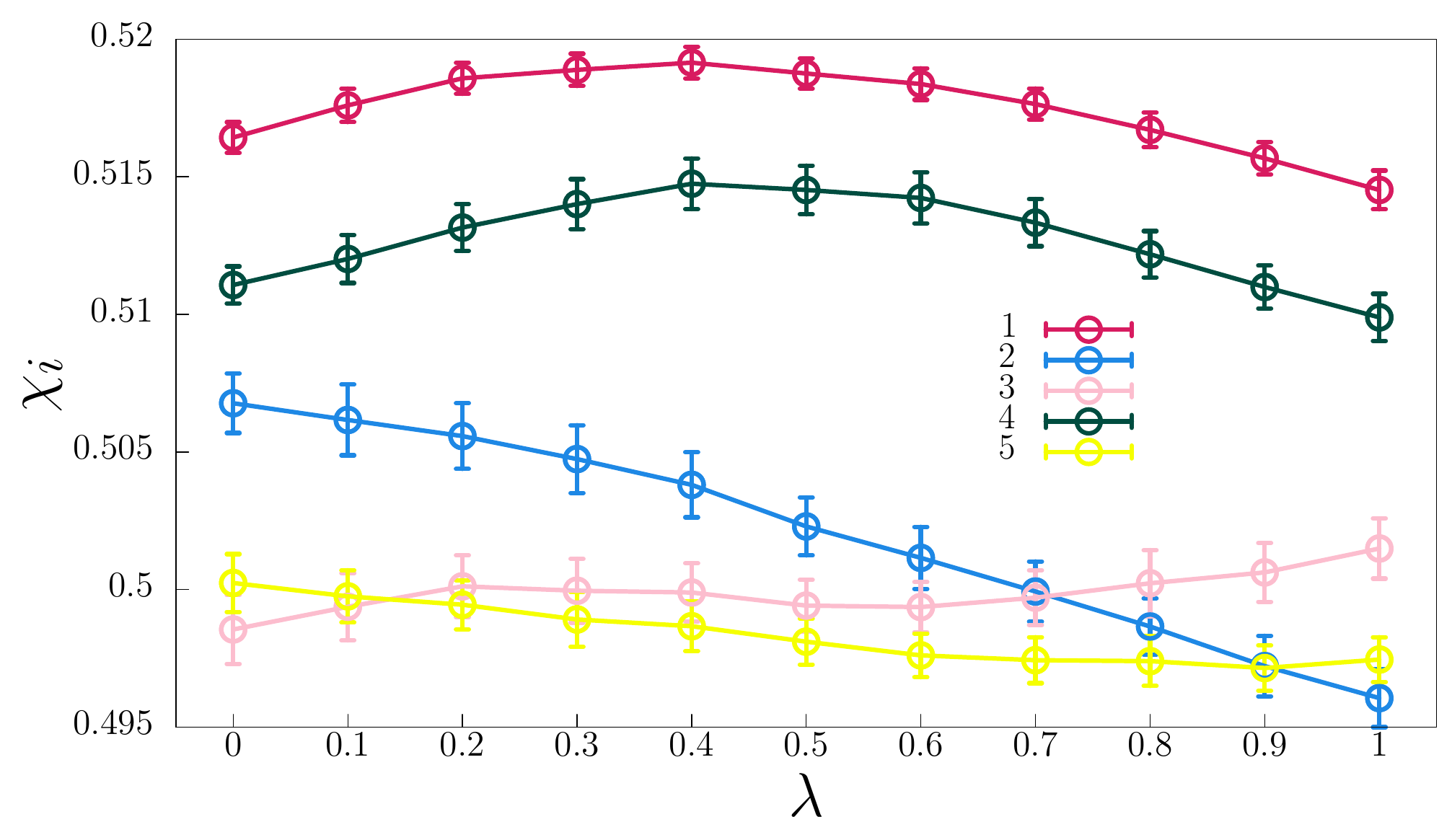}
        \caption{}\label{fig7a}
    \end{subfigure} %
       \begin{subfigure}{.47\textwidth}
        \centering
        \includegraphics[width=73mm]{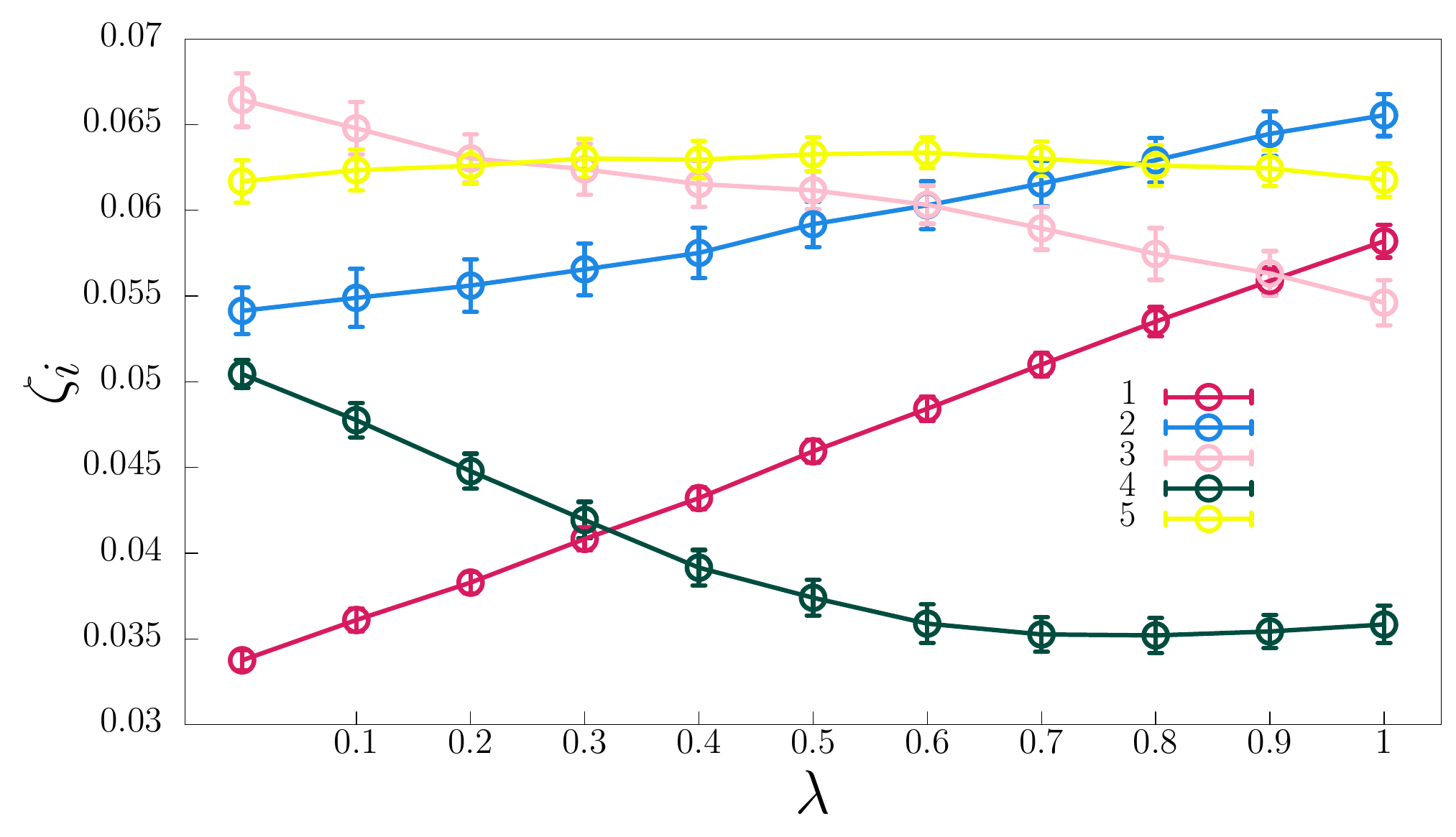}
        \caption{}\label{fig7b}
    \end{subfigure} \\
    \centering
           \begin{subfigure}{.47\textwidth}
        \centering
        \includegraphics[width=73mm]{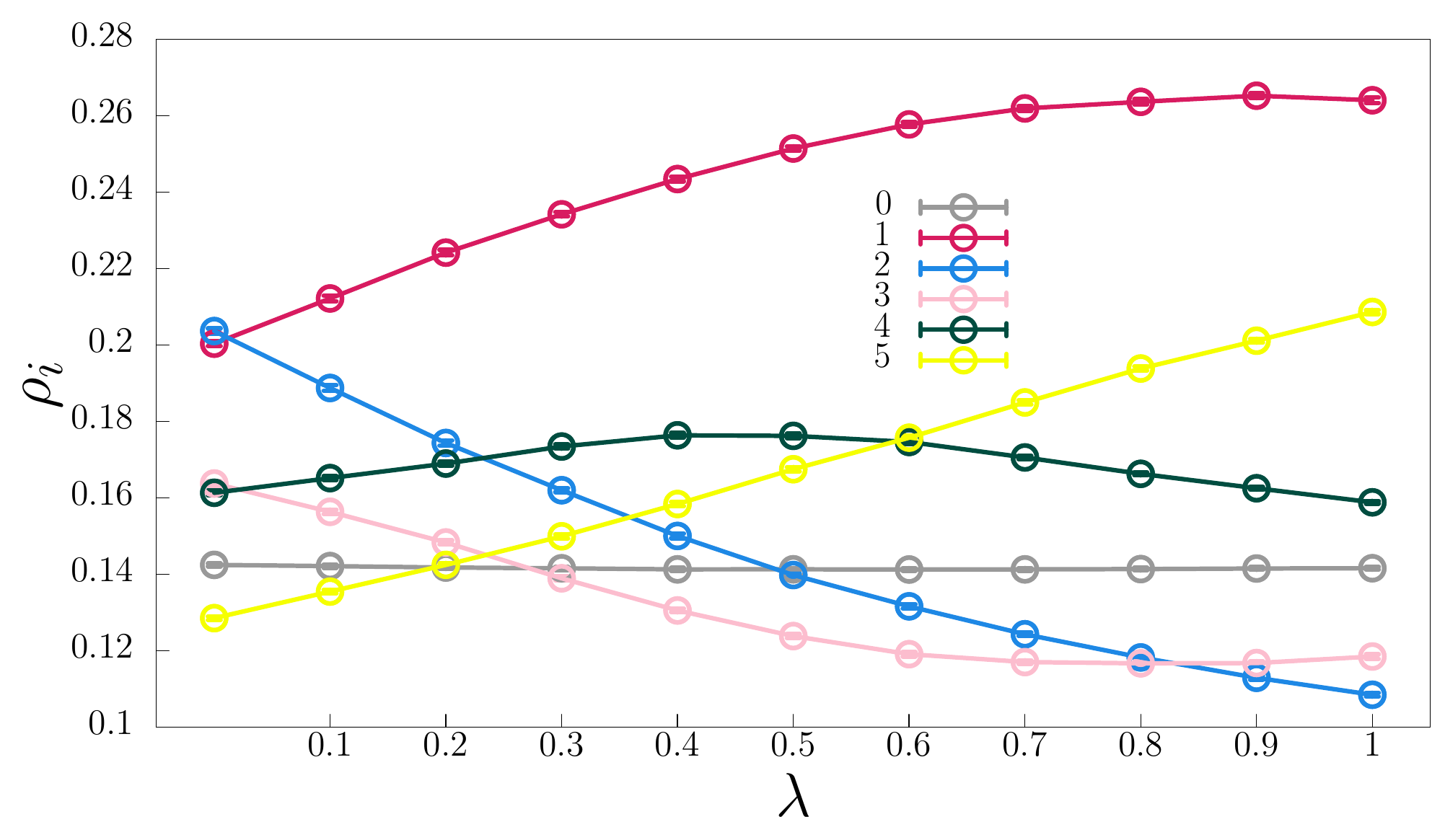}
        \caption{}\label{fig7c}
    \end{subfigure} %
    \caption{Infection risk (Figure a), selection risk (Figure b), and densities of species (Figure c) as functions of the organisms' adaptation factor.
The results were averaged from $100$ simulations for $\beta=0.6$ and $R=3$; the error bars indicate the standard deviation.}
  \label{fig7}
\end{figure}
\section{The impact of the organisms' adaptation factor}
\label{sec5}
Finally, we study how the organisms' conditioning to interpret the environmental cues and react correctly to a local disease surge 
controls the benefits of the locally adaptive survival movement strategy. For this, we ran sets of $100$ simulations for many values $\lambda$, as depicted in Fig.~\ref{fig7a} (infection risk), 
\ref{fig7b} (selection risk), and \ref{fig7c} (density of species).
Following the outcomes in Fig.~\ref{fig5c}, we have chosen $\beta=0.6$ and $R=3$ that maximises the benefits of territorial control for species $1$. 

Fig.~\ref{fig7a} shows that organisms of species $1$ are riskier of being infected than the others. This happens because only organisms of species $1$ perform the Social Distancing tactic, approaching individuals of species $4$, while organisms of other species move randomly.
The outcomes show that if the fraction of organisms of species $1$ conditioned to perceive the presence of a local epidemic outbreak and start the Social Distancing tactic is small, $\chi_1$ grows instead of decreasing. This counterintuitive result holds for $\lambda \leq 0.7$; however, if more than $70\%$ of individuals learn to identify local surges, the infection risk of organisms of species $1$ decreases. The results show that, for $\lambda >0$, organisms of species $2$ and $3$ also benefit from dropped infection risk - although they do not perform any survival movement strategy.

Regarding the selection risk, the results depicted in Fig.~\ref{fig7b} show that the risk of organisms of species $1$ being killed in the cyclic game rises as $\lambda$ grows. This happens because as the proportion of individuals of species $1$ learning to avoid social grows, the vulnerability to being caught by an enemy increases. However, Fig.~\ref{fig7c} shows that the balance is positive for species $1$, which occupies the greater fraction of the grid as $\lambda$ increases. The results show that the benefits for species $1$ are maximised if $90\%$ of individuals are conditioned to adjust the survival movement strategy in response to local epidemic outbreaks. 

Fig.~\ref{fig7c} reveals that as the proportion of organisms of species $1$ learning to change the survival strategy when facing a local disease surge grows, the species $5$ also profits more with population growth. In the case of species $4$, the spatial density increases for $\lambda \leq 0.8$, with the best results obtained if $\lambda \approx 0.4$. In  contrast, the population decline of species $2$ and $3$ increase as $\lambda$ grows;
the density of empty remains approximately constant with the adaptation factor.

\section{Discussion and Conclusions}
\label{sec6}

We explored a generalised rock-paper-scissors model with a disease being transmitted between neighbours, irrespective of the species. We consider that organisms of one out of the species execute a behavioural movement strategy, where each organism moves in the direction with a higher density of individuals that can protect them against enemies. However, the alliance with enemies of enemies is not advantageous when a local surge represents an increased infection risk. In this case, if an organism is apt to recognise the existence of a local epidemic outbreak, the Safeguard strategy is avoided. Instead, the Social Distancing tactic is performed, with the individual moving in the direction with the highest density of empty spaces. In contrast with the model investigated in Ref.~\cite{combination}, where a fixed global design for the combination of survival strategies is assumed by any organism (independent of what is happening in its neighbourhood), here, the choice of the movement tactic is individual. Therefore, each organism is autonomous in taking the more appropriate decision according to the current circunstances, with the Social Distancing tactic substituting the Safeguard strategy only when the local epidemic outbreak reaches the organism's neighbourhood.

The unevenness introduced by the behavioural movement strategy impacts the pattern formation, with the social distancing trigger determining the scales of the spatial regions inhabited by each species and controlling the level of protection against selection and infection. Our findings show a crossover between a regime where organisms focus only on the Social Distancing strategy (low social distancing trigger) and a scenario where they are exclusively interested in protecting themselves against death in the cyclic game (high social distancing trigger). We found that the profit in terms of population growth is maximised if both survival strategies are executed, with organisms using the Social Distancing strategy only when a local surge appears in the neighbourhood.

The organisms' perception radius plays a central role in the local adaptation of the behavioural strategy. The results show that the further the organisms can scan the vicinity, the easier the detection of a forthcoming disease surge and  
more accurate is the decision of the best direction to move.
This means a long-range perception radius produces a more significant reduction in the risks of individuals being killed by an enemy and contaminated by disease. The consequence is the rise of the species population for a large perception radius. 

Our outcomes also show that the organisms' adaptation factor influences population dynamics. We found that,
independent of the proportion of individuals conditioned to move adaptively, the species prevails in the cyclic game.
However, the fraction of territory controlled by the species increases as the adaptation factor increases.

Our simulations assume that each individual can distinguish sick from healthy neighbours; for example, symptoms of the disease are identified as a result of the infection. However, we can extend our conclusions for the case where neither all ill organisms present symptoms. As the individuals are not fully aware of the local surges, the level of precaution should be increased. Suppose that the level of symptomatic disease is $\eta$, where $\eta$ is a real number ($0\,\leq\,\eta\,\leq\,1$). In this case, one has $\beta = \eta\,\tilde{\beta}$, where $\tilde{\beta}$ is the minimum local density of ill organisms with symptoms triggering the Social Distancing tactic.
\section*{Acknowledgments}
We thank CNPq, ECT, Fapern, and IBED for financial and technical support.

\section*{References}
\bibliographystyle{iopart-num}
\bibliography{ref}
\end{document}